\documentclass[usenatbib]{mn2e}

\usepackage{times}
\usepackage[dvips]{graphicx}

\newcommand{\de}{\mathrm{d}} 
\newcommand{\ind}{\indent}
\newcommand{\hmpc}{\ensuremath{h^{-1}\,\hbox{Mpc}}}
\newcommand{\etal}{{et~al.}}
\newcommand{\bfg}{{\bf g}}
\newcommand{\bfr}{{\bf r}}
\newcommand{\bfv}{{\bf v}}
\newcommand{\bfx}{{\bf x}}
\newcommand{\eps}{{\epsilon}}
\newcommand{\lan}{\langle}
\newcommand{\ran}{\rangle}

\renewcommand{\labelenumi}{\Alph{enumi}.}

\title[Velocity--density relation]{The velocity--density relation in the spherical model}

\author[Bilicki \& Chodorowski]{Maciej Bilicki\thanks{E-mail:
    bilicki@camk.edu.pl} and Micha{\l} J.\ Chodorowski\thanks{E-mail:
    michal@camk.edu.pl} \\ N. Copernicus Astronomical Center, Bartycka
  18, 00--716 Warsaw, Poland}

\begin{document}
\maketitle
\begin{abstract}
We study the cosmic velocity--density relation using the spherical
collapse model (SCM) as a proxy to non-linear dynamics. Although the
dependence of this relation on cosmological parameters 
is known to be weak, we retain the density parameter $\Omega_{\rm m}$ in SCM equations, in order to study the limit $\Omega_{\rm m} \to 0$. We show that in this regime the considered relation is strictly 
linear, for arbitrary values of the density contrast, on the contrary to some claims in the literature. On the other hand, we confirm that for 
realistic values of $\Omega_{\rm m}$ the exact 
relation in the SCM is well approximated by the classic formula of Bernardeau (1992), both for voids ($\delta<0$) and for overdensities up to \mbox{$\delta 
\sim 2$ -- $3$.} Inspired by this fact, we find further analytic approximations to the relation for the 
whole range $\delta \in [-1, \infty)$. Our formula for voids 
accounts for the weak $\Omega_{\rm m}$-dependence of their maximal rate of 
expansion, which for $\Omega_{\rm m} < 1$ is slightly smaller that $3/2$. For 
positive density contrasts, we find a simple relation
\begin{displaymath}
\nabla \cdot \mathbf{v} = 3 H_0 \, \Omega_{\rm m}^{0.6}
\left[(1+\delta)^{1/6}-(1+\delta)^{1/2}\right] ,
\end{displaymath}
that works very well up to the turn-around (i.e. up to $\delta \la 13.5$ for $\Omega_{\rm m} = 0.25$ and neglected $\Omega_{\Lambda}$). 
Having the same second-order expansion as the formula of Bernardeau, 
it can be regarded as an extension of the latter for higher density
contrasts. Moreover, it gives a better fit to results of cosmological numerical simulations. 
\end{abstract}


%%%%%%%%%%%%%%%%%%%%%%%%%%%%%%%%%%%%%%%%%%%%%%%%%%%%%%%%%%%%%%%%%%%%%%%%%%%%%%%%%%
\section{Introduction}
\label{sec:intr}
The gravitational instability is commonly accepted as the process of
large-scale structure formation in the Universe. According to this
scenario, structures formed by the growth of small
inhomogeneities present in the early Universe.  Gravitational
instability gives rise to a coupling between the density and peculiar
velocity fields of matter. On very large, linear scales, the relation between
the peculiar velocity $\mathbf{v}$ and the density contrast $\delta$
in co-moving coordinates is
\begin{equation}
\nabla \cdot \mathbf{v}(\mathbf{x}) = - H f(\Omega,\Lambda) \,
\delta(\mathbf{x}) \,,
\label{eq:linear}
\end{equation}
where $H$ is the Hubble constant. [For simplicity of notation, we use the notation ($\Omega$, $\Lambda$) instead of ($\Omega_\mathrm{m}$, $\Omega_\Lambda$).] The coupling constant, $f$, carries
information about the underlying cosmological model and is related to
the cosmological matter density parameter, $\Omega$, and cosmological
constant, $\Lambda$, by
\begin{equation}
f(\Omega,\Lambda) \simeq \Omega^{0.6} +
\frac{\Lambda}{70} \left(1 + \frac{\Omega}{2}\right)
\label{eq:f_Omega_Lambda}
\end{equation} 
\citep{LLPR}. The linear amplitude of peculiar velocities is
thus sensitive to $\Omega$; on the other hand, it is quite insensitive
to~$\Lambda$. Hence, comparing the observed density and velocity
fields of galaxies allows one to constrain $\Omega$, or the degenerate
combination $\beta \equiv \Omega^{0.6}/b$ in the presence of so called galaxy
biasing (e.g. \citealt{SW} for a review). This is done by
extracting the density field from all-sky redshift surveys -- such as
the Point Source Catalogue Redshift survey \citep[PSCz,][]{Saund}, or the 2MASS Redshift Survey \citep[2MRS,][]{Huch} -- and
comparing it with the observed velocity field from peculiar velocity
surveys. The methods for doing this fall into two broad
ca\-te\-go\-ries. One can use Equation~(\ref{eq:linear}), calculating the
divergence of the observed velocity field and comparing it directly with the
density field from a redshift survey; this is referred to as a
\textit{density--density comparison}. Alternatively, one can use the integral
form of Equation~(\ref{eq:linear}) to calculate the predicted
velocity field from a redshift survey, and compare the result with the
measured peculiar velocity field; this is called a \textit{velocity--velocity
comparison}. Velocity--velocity comparisons are generally regarded as
more reliable, since they involve manipulation of the denser and more
homogeneous redshift catalogue data, while density--density
comparisons require manipulation of the noisier and sparser velocity
data. In both cases, the density and velocity fields need to be
smoothed in order to reduce errors and shot noise. Velocity--velocity
comparisons require a smaller size of smoothing, of a few \hmpc. For example,
\citet{Will} used a smoothing scale of $3$ \hmpc. Such
scales are called \textit{mildly non-linear}: the variance of the density field
smoothed over the scale of a few \hmpc\ is of order unity.

Mildly non-linear extensions of Equation~(\ref{eq:linear}) have been
developed by a number of workers. These extensions have been based
either on various analytical approximations of non-linear dy\-na\-mics
(\citealt{ReGe}; Bernardeau 1992, hereafter \citealt{B92}; \citealt{Cat}; \citealt{Ch97}; \citealt{ChLo}; \citealt{Ch98}), or numerical (either
N-body or hydrodynamic) simulations (\citealt{Man}; Kudlicki \etal 2000, hereafter \citealt{KaCPeR}),
or both (\citealt{NDBB}; \citealt{Gram}; \citealt{MY};
Bernardeau \etal 1999, hereafter \citealt{B99}). Unlike the linear case~(\ref{eq:linear}), the non-linear relation between the velocity divergence and
the density contrast at a given point is non-deterministic (though in
the non-linear regime the two fields remain highly
correlated). Therefore, for a full description of the relation, the
conditional means (mean $\nabla \cdot \bmath{v}$ given $\delta$ and
vice versa) are not sufficient: one has to describe the full bivariate
distribution function for $\nabla \cdot \bmath{v}$ and $\delta$, or at
least the conditional scatter. These aspects of the velocity--density
relation were studied by \citet{Ch98} and more
extensively by \citet{B99}. However, in practical
applications the intrinsic scatter in the velocity--density relation
is much smaller than the one induced by observational errors, and the
conditional means are sufficient.

\citet{B99} and \citet{KaCPeR} found that very good fits to
the mean relations, obtained for the mildly non-linear fields extracted
from numerical simulations, were given by modifications of the formula
of \citet{B92}. This formula describes a non-linear relation between
initially Gaussian, random fields of $\nabla \cdot \bmath{v}$ and
$\delta$, under the assumption of a vanishing variance of the density
field (so the relation has no scatter). \citet{B92} claimed his
relation to be the same as the one exhibited in the spherical collapse
model (hereafter SCM). In practical applications (namely with non-zero
variance of the density field), he predicted his formula to work well
in voids, but `to become very inaccurate for $\delta$ larger than 1 or
2'.

In {\em this\/} paper we study the velocity--density relation in the
SCM. The reason for such an approach is twofold. First, to derive his
formula, \citet{B92} used quite sophisticated methods (summing up first
non-vanishing contributions from the reduced part of {\em all-order\/}
joint moments of $\nabla \cdot \bmath{v}$ and $\delta$). On the other hand, the dynamics
of the SCM is very simple and should allow to re-derive the formula of
\citet{B92} in a straightforward way. More importantly, in the SCM the
relation can be easily extended to higher values of $\delta$, with the
hope that this modification will fit better the results of numerical
experiments of \citet{B99} and \citet{KaCPeR}. The SCM is in
principle insensitive to the variance of the density field (and the
resulting velocity--density relation is deterministic), but in
practice the variance of the smoothed density field dictates how high
density contrasts can be reached.

The non-linear relation between $\delta$ and $f^{-1} \nabla \cdot
\mathbf{v}$ (note the scaling $f^{-1}$) depends very weakly on cosmological parameters. \citet{B92} analysed the
$\Omega$-dependence of the scaled velocity--density relation in the
limit $\lan\delta^2\ran \to 0$ and found it to be very weak. \citet{Bouch} showed that second and third order expansions for
$\delta$ and $f^{-1} \nabla \cdot \mathbf{v}$ depend extremely weakly
on $\Omega$ and $\Lambda$. \citet*{SCF}
demonstrated that this is the case for {\em all\/} orders. Specifically,
they showed that perturbative solutions for the density contrast for
arbitrary cosmology are, with a good accuracy, separable: $\delta_n =
D^n(t)\, \eps_n(\bfx)$, where $D(t)$ is the linear growing mode for
this cosmology and $\eps_n$ is the spatial part of the $n$-th order
solution for the Einstein--de Sitter model. Using the continuity
equation one can then prove, by induction, that the velocity
divergence depends on $\Omega$ and $\Lambda$ practically only through
the factor $f(\Omega,\Lambda)$. Most generally, Nusser \& Colberg (1998) (hereafter \citet{NuCo}) showed {\em the equations of motion\/} of the cosmic
pressureless fluid to be `almost independent' of cosmological
parameters. The weak dependence of the scaled velocity--density
relation on the background cosmological model has been also confirmed
by N-body numerical simulations (\citealt{Man}; \citealt{B99}).

However, the $\Omega$-dependence of the equations of motion of the
cosmic dust stops to be weak when $\Omega \ll 1$ (see eqs.~13--14 of
\citealt{NuCo}). This regime of $\Omega$ is not physically
relevant, since the currently preferred value of $\Omega$ is much
higher. Still, \citet{B92} derived his formula applying the limit $\Omega \to
0$. Therefore, in the present paper we will neglect $\Lambda$ (setting $\Lambda = 0$), but will retain the $\Omega$-dependence of the
equations of the spherical collapse and in particular examine the limit of small $\Omega$.

The paper is organised as follows. Section \ref{sec:csm} presents
general assumptions, terminology and basic formulae of the spherical model. In
Section \ref{sec:pp} we focus on the factor $f$, appearing in
Eq.~(\ref{eq:linear}) and commonly approximated by
Formula~(\ref{eq:f_Omega_Lambda}), or its simplified version $f\simeq\Omega^{0.6}$ . Section \ref{sec:smallOm} contains
an analysis of the regime of very small $\Omega$ and presents the
resulting universal velocity--density relation. In Sections \ref{sec:voids} and \ref{sec:overd}, basing on analytical considerations, we derive approximations for the
relation between the velocity divergence and the density contrast
respectively for spherical voids and overdensities, for realistic
values of $\Omega$. These approximations constitute the main results
of this paper. Section \ref{sec:comp} gives a comparison of our fits with results of numerical simulations. We conclude in Section \ref{sec:concl}.

%%%%%%%%%%%%%%%%%%%%%%%%%%%%%%%%%%%%%%%%%%%%%%%%%%%%%%%%%%%%%%%%%%%%%%%%%%%
\section{Cosmological spherical model}
\label{sec:csm}

\ind Let us consider an open Friedman world model (i.e.\ with
$\Omega<1$) without the cosmological constant, $\Lambda=0$. We
introduce the \textit{conformal time} $\eta$ related to the cosmic
time $t$ by the equation
\begin{equation}\label{form:conft}
\de\eta=\frac{c\,\de t}{R_0\,a}\,,
\end{equation}
where $R_0=c\slash(H_0 \sqrt{1-\Omega_0})$ is the curvature radius of
the universe, $c$ and $a$ are respectively the velocity of light and
the scale factor; subscripts `0' and (used later) `i' refer
to the present day and to some adequately chosen initial moment,
respectively. Now, the time evolution of the scale factor can be expressed
in terms of the following parametric equations (e.g.\ \citealt{Pe80}):
\begin{equation}\label{eq:a_t_eta}
a(\eta)=A\left(\cosh\eta-1 \right) ,\quad t(\eta)=B\left(
\sinh\eta-\eta\right)\quad(\eta\ge0) \,,
\end{equation}
where $A$ and $B$ are constants. Moreover, in this model the conformal
time $\eta$ is unambiguously related to the density parameter $\Omega$
by
\begin{equation}\label{form:Omega of eta}
\Omega=\frac{2}{1+\cosh\eta}\,.
\end{equation}

If we now consider a {\em top-hat\/} spherical perturbation (a sphere
of homogeneous density embedded in a Friedman universe), it can be
analysed as a `universe of its own' (as was noted for the first time
by \citealt{Lem}) with a scale factor $a_\mathrm{p}$ which is the
radius of the perturbation. Introducing the \textit{density contrast}
of the perturbation relative to the background, $\delta$, as
\begin{equation}
\delta\equiv\frac{\rho_\mathrm{p}-\rho_\mathrm{b}}{\rho_\mathrm{b}}\,,
\end{equation}
we obtain two cases to be taken into account. Using the same
terminology for spherical perturbations as for analogous Friedman
world models, an \textit{open} perturbation is such that its initial
density contrast $\delta_\mathrm{i}$ is smaller than the
\textit{critical density contrast} $\delta_\mathrm{c}$ (the density contrast of an
Einstein--de Sitter type of perturbation, i.e. with
$\Omega^{(\mathrm{p})}=1$), given by:
\begin{equation}\label{form:delta_c}
\delta_\mathrm{c}\equiv \frac{3}{5} \left(\Omega^{-1}_\mathrm{i}-1\right) .
\end{equation}
It can be checked that the density parameter of thus defined open perturbation
is $\Omega^{(\mathrm{p})}<1$, as expected. These results are valid
under the assumption that the initial density of the background is
sufficiently close to the critical density
($\Omega_\mathrm{i}\simeq1$). The factor $3/5$ in Eq. (\ref{form:delta_c}) comes from the decomposition of the density field into two
components, one related to the growing mode and the other to the
decaying one; we assume here the perturbation to be purely in the
growing mode. For details see \citet{Pe80}.

The evolution of such a spherical perturbation is governed by
equations analogous to (\ref{eq:a_t_eta}):
\begin{equation} \label{eq:a_t_phi}
a_\mathrm{p}(\phi)=A_\mathrm{p} \!\left(\cosh\phi-1 \right)\!,~~
t(\phi)=B_\mathrm{p} \!\left( \sinh\phi-\phi\right) ~~ (\phi\ge0) . 
\end{equation} 
The normalization factors are such that \mbox{$(A_\mathrm{p}\slash
A)^3=(B_\mathrm{p}\slash B)^2$}. Of course, time $t$ is the same for
the background as for the perturbation, which leads to the relation
between $\phi$ and $\eta$:
\begin{equation}\label{eq:sinh fi fi}
\sinh\phi-\phi=(1-r)^{3\slash 2}(\sinh\eta-\eta)\,,
\end{equation}
where we have used $r\equiv\delta_\mathrm{i}\slash\delta_\mathrm{c}$
(for a detailed derivation see \citealt{Pe80}). If $\delta>0$ then $r
> 0$, so $\phi<\eta$, and vice versa for negative $\delta$.

In order to obtain similar relations for a \textit{closed}
perturbation ($\delta_\mathrm{i}>\delta_\mathrm{c}$ or
$\Omega^{(\mathrm{p})}>1$), one should make the following
substitutions:
\begin{equation}\label{substitutions}
\phi\rightarrow i\phi\,,\qquad A_\mathrm{p} \rightarrow
-A_\mathrm{p}\,, \qquad B_\mathrm{p} \rightarrow iB_\mathrm{p}\,,
\end{equation}
remembering that in such a case $0\le\phi\le2\pi$.

We can now express the density contrast in terms of the parameters
$\eta$ and $\phi$, using the relation $\rho_\mathrm{p}
\slash\rho_\mathrm{b} =(a\slash a_\mathrm{p})^3$:
\begin{equation}\label{form:delta fi eta}
\delta=\left(\frac{\sinh\phi-\phi}{\sinh\eta-\eta}\right)^2 
\left(\frac{\cosh\eta-1}{\cosh\phi-1}\right)^3 - 1
\end{equation}
for open perturbations and
similarly for closed ones, with the use of (\ref{substitutions}):
\begin{equation}\label{form:delta fi eta closed}
\delta=\left(\frac{\phi-\sin\phi}{\sinh\eta-\eta}\right)^2 
\left(\frac{\cosh\eta-1}{1-\cos\phi}\right)^3 - 1
\end{equation}
(cf. \citealt{ReGe}; \citealt{FoGa}). Note
that always $\delta\ge-1$, but in principle the density contrast has
no upper bound. However, if initially
$0<\delta_\mathrm{i}<\delta_\mathrm{c}$, then $\delta$ cannot exceed a
maximal value which can be calculated taking $\phi\to0$ in
(\ref{form:delta fi eta}):
\begin{equation}\label{form:delta lim}
\delta_\mathrm{lim} = 
\frac{2}{9}\frac{(\cosh\eta-1)^3}{(\sinh\eta-\eta)^2}-1\,.
\end{equation}
The above value becomes the minimal value of the density contrast for
closed perturbations, i.e. it is a boundary value of possible density contrasts
between closed and open perturbations for a given $\eta$.

The linear theory relates the density contrast of a perturbation to
its \textit{peculiar velocity divergence} $\nabla\cdot\mathbf{v}$
(Eq. \ref{eq:linear}). In the spherical model we obtain
$\nabla\cdot\mathbf{v}=3(H_\mathrm{p}-H)$, where
$H_\mathrm{p}=\dot{a}_\mathrm{p}\slash a_\mathrm{p}$. For convenience
we change units and sign, obtaining what will be called in this paper
the (dimensionless) velocity divergence, $\theta$:
\begin{equation}
\theta=3\left(1-\frac{H_\mathrm{p}}{H}\right) .
\end{equation}
Some simple algebra is sufficient to find the dependence of
\mbox{$H_\mathrm{p}\slash H$} on $\eta$ and $\phi$, which leads to the
following expression (\citealt{ReGe}; \citealt{B99}):
\begin{equation}\label{form:teta fi eta}
\theta=3\left[1-\frac{\sinh\phi\,(\sinh\phi-\phi)}{\sinh\eta\,(\sinh\eta-\eta)}
\left(\frac{\cosh\eta-1}{\cosh\phi-1}\right)^2\right] ,
\end{equation}
valid for open perturbations on an open background; substitution $\phi
\rightarrow i\phi$ gives the relation for closed perturbations:
\begin{equation}\label{form:teta fi eta closed}
\theta=3\left[1-\frac{\sin\phi\,(\phi-\sin\phi)}{\sinh\eta\,(\sinh\eta-\eta)}
\left(\frac{\cosh\eta-1}{1-\cos\phi}\right)^2\right] .
\end{equation}
Both the density contrast and the velocity divergence, as given by (\ref{form:delta fi eta}) and (\ref{form:teta fi eta}), or (\ref{form:delta fi eta closed}) and (\ref{form:teta fi eta closed}), are parametrically
dependent on $\phi$ ($\eta$ is fixed). Our aim here is to eliminate
this parameter (at least approximately) and to obtain the
$\theta$--$\delta$ relation in the spherical model in an analytic
form. 

As a first step, it is useful to simplify the formula for $\theta$
including `the easy part' of the dependence on $\delta$. This is
done by calculating \mbox{$(\sinh\phi-\phi)\slash(\sinh\eta-\eta)$} from (\ref{form:delta fi eta}) and inserting the resultant
expression into
(\ref{form:teta fi eta}). Then, owing to the hyperbolic identity
$\cosh^2x-\sinh^2x=1$ and the relation (\ref{form:Omega of eta}) for
$\Omega$, we finally obtain a simplified formula for the velocity
divergence:
\begin{equation}\label{form:teta open}
\theta=3\left[1-\sqrt{\frac{1}{2}\Omega(1+\delta)(1+\cosh\phi)}\right].
\end{equation}
These considerations were valid for open perturbations. If
\mbox{$\Omega^{(p)}>1$,} then we have
\begin{equation}\label{form:teta closed}
\theta=3\left[1\mp\sqrt{\frac{1}{2}\Omega(1+\delta)(1+\cos\phi)}\right],
\end{equation}
where `--' applies to the case $0\le\phi<\pi$ and `+' to $\pi \le
\phi \le 2\pi$. Formula (\ref{form:teta open}) [(\ref{form:teta closed})] is simpler than (\ref{form:teta fi eta}) [(\ref{form:teta fi eta closed})], but the dependence on $\phi$ remains; the parameter $\phi$ is related to $\delta$ by Equation~(\ref{form:delta fi eta}) [(\ref{form:delta fi eta closed})].

%%%%%%%%%%%%%%%%%%%%%%%%%%%%%%%%%%%%%%%%%%%%%%%%%%%%%%%%%%%%%%%%%%%%%%%%%%
\section{Factor \lowercase{$\bmath{f}$}}
\label{sec:pp}

\ind The linear theory (valid for small values of $\delta$) relates
the velocity divergence as defined above to the density contrast
through the equation
\begin{equation}\label{eq:lin_the_f_del}
\theta=f\delta 
\end{equation}
[cf.\ Eq.~(\ref{eq:linear})], where the factor $f=f(\Omega,\Lambda)$ is
given by
\begin{equation}
f\equiv\frac{\de\ln D}{\de\ln a}\,.
\end{equation}
The quantity $D(t)$ is the growing mode of the perturbation. The factor
$f$ has been a subject of study in many papers (e.g. \citealt{Pe76};
\citealt{LiSch}; \citealt{LLPR}; \citealt{Mar}; \citealt{Bouch};
\citealt{FoGa}; \citealt{NuCo}). The best-known and most widely used
approximation (often without reference) is the one given by
\citet{Pe76}:
\begin{equation}\label{form:Peebles}
f(\Omega)\simeq\Omega^{0.6}\,.
\end{equation}
In this part we will compare this fit with the exact formula for $f$.

\begin{figure}
\begin{center}
\includegraphics[angle=270, width=0.5\textwidth]{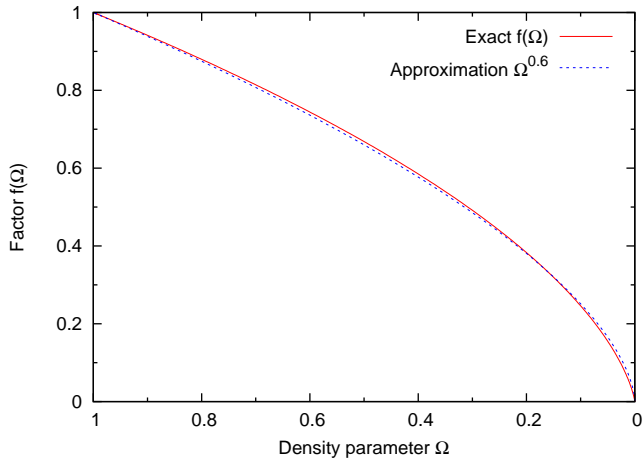} 
\end{center}
\caption{\label{fig:fofOmega} Factor
  $f\equiv\left(\de\ln D\right)\slash\left(\de\ln
  a\right)$ as a function of density parameter $\Omega$ for world
  models with $\Lambda=0$: exact relation (solid line) and
  approximation $f\simeq\Omega^{0.6}$ (dashed line).}
\end{figure}

The spherical model as described here allows us to calculate
$f(\Omega,\Lambda=0)$ as the limit
\begin{equation}
f=\lim_{\delta\to 0}\frac{\theta}{\delta}\,.
\end{equation}
It can be checked that choosing $|\delta| \ll 1$ is equivalent to
taking $|r| \ll 1$ [Eq. (\ref{eq:sinh fi fi})]. Moreover, from the
relation (\ref{eq:sinh fi fi}) it follows that in this case
$\phi=\eta+\varepsilon$, where $|\varepsilon| \ll 1$. Using the
first-order approximation $(1-r)^{3\slash2}\simeq1-\frac{3}{2} r$ and
expanding hyperbolic functions around $\varepsilon=0$, we can
linearize Equations (\ref{eq:sinh fi fi}), (\ref{form:delta fi eta})
and (\ref{form:teta fi eta}). As a result we get a linear relation
between $\varepsilon$ and $r$ and further on also linear dependencies
of $\delta$ and $\theta$ on $r$. Diving thus obtained velocity
divergence by the density contrast, we get the following formula for
$f$ as a function of $\eta$:
\begin{equation}\label{form:f of eta}
f(\eta)=\frac{3\,\eta\, (2+\cosh\eta)-9\sinh\eta}{3\, (\cosh\eta+1)\, 
(\sinh\eta-\eta) - 2\sinh\eta\, (\cosh\eta-1)} \,.
\end{equation}
A similar relation, but for a `closed' model of the background, can be
found in \citet{LiSch}. If we now make the substitution
$\eta=\mathrm{arcosh}\left({2}\slash\Omega-1\right)$ [Eq. (\ref{form:Omega of eta})], then after some
algebra we can express the parameter $f$ as a function of $\Omega$:
\begin{displaymath}
f(\Omega)= 
\end{displaymath}
\begin{equation}
\frac{3\, \Omega\, (\Omega+2)\ln[2\Omega^{-1}(1+\sqrt{1-\Omega})-1]
  - 18\, \Omega\, \sqrt{1-\Omega}}{12\sqrt{1-\Omega}-8\sqrt{(1-\Omega)^3} - 
6\Omega\ln[2\Omega^{-1}(1+\sqrt{1-\Omega})-1]}.
\label{form:f of Omega}
\end{equation}
This is the exact expression for $f(\Omega)$ with $\Lambda=0$ and
$\Omega<1$. It was already derived for example by
\citet{FoGa}. Figure~\ref{fig:fofOmega} presents a comparison of this
relation with the Peebles' formula $\Omega^{0.6}$. It can be seen that
the power-law approximation is sufficiently exact, especially for the
currently favoured value of the density parameter ($\Omega\simeq
0.25$). Moreover, owing to the complicated form of (\ref{form:f of Omega}), the latter is not very useful. However, one should always
bear in mind that the formula (\ref{form:Peebles}) is merely an
approximation and in some applications its usage may lead to errors. A
much better fit is the one given in a footnote of \citet{NuCo}:
$f=\Omega^{4/7+(1-\Omega)^3\!/20}$. Its errors relative to the exact
value for the model with $\Lambda=0$ are below 0.3\% for $\Omega>0.1$.

%%%%%%%%%%%%%%%%%%%%%%%%%%%%%%%%%%%%%%%%%%%%%%%%%%%%%%%%%%%%%%%%%%%%%%%%%%%%%%%%%%%%%%%%%
\section{Limit of small $\mathbf \Omega$}
\label{sec:smallOm}

\begin{figure}
\begin{center}
\includegraphics[angle=270, width=0.5\textwidth]{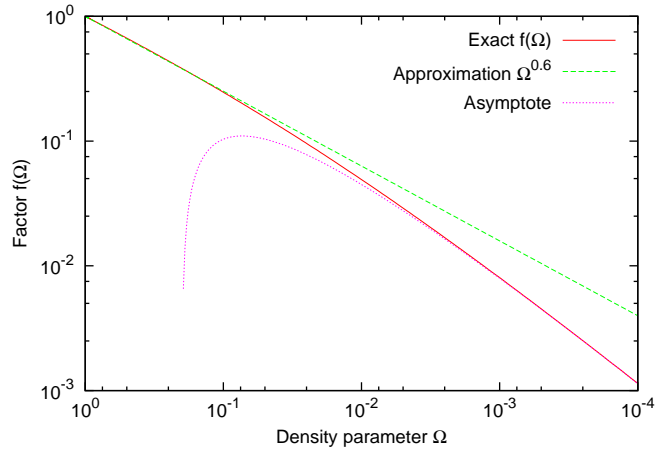} 
\end{center}

\caption{\label{fig:f of Om small} Factor $f(\Omega)$
in the limit $\Omega\ll 1$: exact relation (solid line), power-law
approximation (dashed line) and asymptotic relation \mbox{$f(\Omega)\simeq-(3\slash2)\Omega(\ln\Omega + 3 - \ln{4})$} -- dotted line. Note logarithmic scales on both axes.}
\end{figure}

Let us now examine more thoroughly the limit of $\Omega\to 0$. We begin with checking the asymptotic behaviour of $f(\Omega)$. This regime, although not physically
interesting, allows to take a closer look on the bottom-right end of the diagram
presented in Fig. \ref{fig:fofOmega}, and the results obtained will be useful later in the paper. (See also Appendix~\ref{app:empty}.) Starting with the relation
(\ref{form:f of eta}) and remembering that the limit of small $\Omega$ means
$\eta\gg 1$, we get the following approximation:
\begin{equation}\label{form:f eta big}
f(\eta)\simeq6\mathrm{e}^{-\eta}(\eta-3)\qquad(\eta\gg1)\,.
\end{equation}
If we now observe that for such $\eta$ we also have $\eta\simeq\ln4 -
\ln\Omega$ and $\mathrm{e}^{\eta}\simeq4\Omega^{-1}$, we obtain an
asymptotic formula for $f(\Omega)$:
\begin{equation}\label{form:f Omega small}
f(\Omega)\simeq-\frac{3}{2}\Omega(\ln\Omega + 3 - \ln{4}) 
\qquad (\Omega \ll 1).
\end{equation}
Figure \ref{fig:f of Om small} clearly shows that for sufficiently
small $\Omega$, i.e. $\Omega<0.01$, the power-law of Peebles could no
longer be used. This plot is also a confirmation that in some cases
the usage of log-log diagrams is well-grounded.

\ind \citet{B92} studied the cosmic statistical relation between the
non-linear density contrast and the velocity divergence,
evolving from Gaussian initial conditions, in the limit of a vanishing
variance of the density field. He found the result `to be very close
to'
\begin{equation}\label{form:Bernard Teta}
\Theta^{(\mathrm{B})}=\frac{3}{2}\left[(1+\delta)^{2\slash
3}-1\right] \,.
\end{equation}
Here, and from now on, the so-called \textit{scaled} velocity divergence, $\Theta$, is
defined as
\begin{equation}\label{form:Teta}
\Theta = f^{-1} (\Omega)\, \theta \,.
\end{equation}
Note that for $|\delta| \ll 1$, the non-linear formula~(\ref{form:Bernard Teta}) correctly reduces to $\Theta^{(\mathrm{B})} =
\delta$, i.e. to the linear-theory relation (\ref{eq:lin_the_f_del}). As already mentioned,
\citet{B92} claimed his relation to be the same as the one exhibited in
the SCM. In turn, \citet{B99} argued that the approximation
(\ref{form:Bernard Teta}) `is strictly valid in the limit
$\Omega\to 0$'. Here we check these statements, applying the
regime $\Omega\ga 0$ to the equations of the SCM.

If $\Omega\ll 1$ then $\eta\gg 1$. Therefore, since for voids ($\delta
< 0$) we have $\phi > \eta$, also $\phi\gg 1$. For overdensities
($\delta > 0$), the limit \mbox{$\eta \to\infty$} applied to Eq. (\ref{form:delta lim}) gives $\delta_\mathrm{lim}\to+\infty$. Thus we can focus only on Formula (\ref{form:delta fi eta}) for $\delta$. From Eq. (\ref{form:delta fi eta}) we see that in order to keep
$\delta$ finite (though arbitrarily large), also $\phi$ should tend to
infinity. In other words, if $\eta\gg 1$, then also $\phi\gg 1$, both
for voids and overdensities. Hence, still from Equation~(\ref{form:delta fi eta}), we get, up to the leading order,
\begin{equation}
\phi_1 = \eta - \ln(1+\delta)\;,
\end{equation}
and up to the second order
\begin{equation}
\phi_2 = \eta - \ln(1+\delta) + \left[ 4(1+\delta)\ln(1+\delta) -
  \delta (4\eta-6) \right] {\rm e}^{-\eta} .
\label{eq:phi2_ome_to_0}
\end{equation}
Equivalently,
\begin{equation}
(1+\delta) \cosh\phi_2 = \cosh\eta\, + \frac{1}{2} \left[ 
4(1+\delta)\ln(1+\delta) - \delta (4\eta-6) \right] \!.
\label{eq:cosh_phi2_ome_to_0}
\end{equation}
Applying this formula in Equation~(\ref{form:teta open}) and using
the large-$\eta$ limit of the function $f(\eta)$ (Eq.~\ref{form:f
  eta big}) we obtain
\begin{equation}\label{form:tetafal2}
\Theta \, \simeq \, \delta-\frac{N(\delta)}{\eta-3} \; \simeq \;
\delta + \frac{N(\delta)}{\ln\Omega+3-\ln 4}\,,
\end{equation}
where
\begin{equation}\label{form:N of delta}
N(\delta)=(\delta+1)\ln(\delta+1)-\delta\,.
\end{equation}
In the limit $\Omega\to 0$ the second term in
Equation~(\ref{form:tetafal2}) vanishes, hence
\begin{equation}\label{eq:linear theory}
\Theta = \delta\,.
\end{equation}
This is exactly the same relation as for the linear regime (where
$|\delta| \ll 1$). However, here the density contrast can have an
arbitrary value. Thus, the formula of \citet{B92}~(\ref{form:Bernard
  Teta}) does {\em not\/} describe the dynamics of perturbations in
the limit $\Omega\to 0$. The relation (\ref{eq:linear theory}), being very simple, is a non-trivial result. When $\Omega$
tends to 0, then also the (\textit{not} scaled) velocity divergence
$\theta\to 0$ (peculiar velocities vanish with diminishing $\Omega$). However, the quantity $\Theta$, as introduced
by Eq. (\ref{form:Teta}), converges to a non-zero value for
$\Omega\to 0$, due to the presence of the factor $f(\Omega)$, approximated by (\ref{form:f Omega small}) for small $\Omega$. The normalisation used here leads to $\Theta=\delta$ in
the linear theory. Why for very small values of $\Omega$ this relation
holds also in the non-linear regime? It turns out that this is a
general result of dynamics in a low-density universe, and does not
rely on any symmetry. The derivation is presented in Appendix \ref{app:empty}.

\begin{figure}
\begin{center}
\includegraphics[angle=270, width=0.5\textwidth]{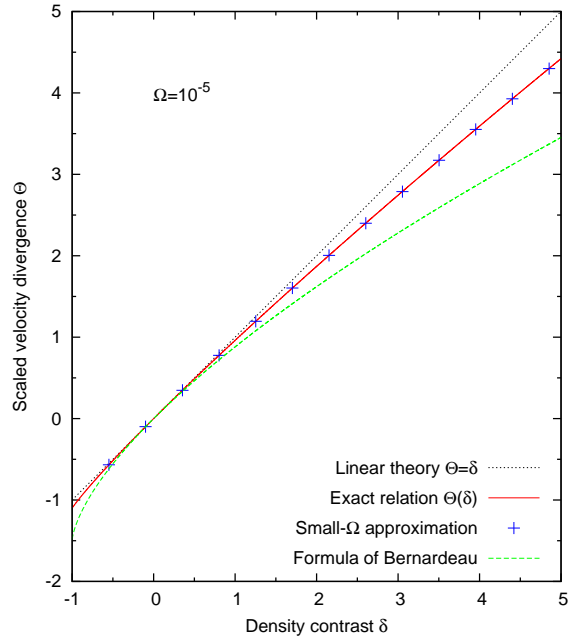} 
\end{center}
\caption{\label{fig:Theta small Om} A comparison of the relation
  between density contrast $\delta$ and scaled velocity divergence
  $\Theta=f^{-1}\,\theta$ for very small $\Omega$ (here
  $\Omega=10^{-5}$):  linear
  theory result $\Theta=\delta$ -- dotted line, exact relation $\Theta(\delta)$ -- solid line,
  approximation for small $\Omega$ -- crosses and \citet{B92}
  approximation (\ref{form:Bernard Teta}) -- dashed line.}

\end{figure}

For large but finite values of $\eta$, Formula~(\ref{form:tetafal2}) can be applied. Specifically, it can be used for $\eta$
significantly greater than~3 ($\Omega$ significantly smaller than~$0.2$), which falls well below the presently accepted value of the
cosmic density parameter. Therefore, the approximation~(\ref{form:tetafal2}) is of no practical importance; we have to continue our
search for a relevant relation. Just for illustrative purposes, on
Figure~\ref{fig:Theta small Om} we plot the exact relation, the
\citet{B92} approximation, and the approximation~(\ref{form:tetafal2}),
for an exemplary value of $\Omega=10^{-5}$ ($\eta\simeq13$). We see
that although for this value of $\Omega$ the relation is still
non-linear, the \citet{B92} approximation drastically overestimates the
degree of non-linearity.

%%%%%%%%%%%%%%%%%%%%%%%%%%%%%%%%%%%%%%%%%%%%%%%%%%%%%%%%%%%%%%%%%%%%%%%%%%%%%%%%%%%%%%%%%
\section{Relations for voids}
\label{sec:voids}

As \textit{voids} we will understand any underdense perturbations,
i.e. those for which $\delta<0$. In this section we examine the
behaviour of the velocity divergence vs. the density contrast for such inhomogeneities.

When considering overdense perturbations (with $\delta>0$), the regime
of $\delta\simeq1$ is usually called \textit{weakly} (or at most
\textit{mildly}) non-linear. It may thus seem that it should be
similarly for the limit $\delta\ga -1$ (cf. \citealt{Mar}). However,
if we analyse Equation (\ref{form:delta fi eta}), which is valid both
for voids and for open overdensities, we can see that for finite
values of $\eta$ (which correspond to non-zero $\Omega$), the
condition $\delta\to -1$ may only be satisfied for
$\phi\to+\infty$. Hence, the evolution of such a perturbation
is \textit{highly} non-linear when the density contrast approaches its minimum
value.

The scaled velocity divergence $\Theta$, as defined in Eq. (\ref{form:Teta}), is a monotonically increasing function of $\delta$ (for
$\eta$, or $\Omega$, treated as a fixed parameter). Its minimum value
is $\Theta_\mathrm{min}\equiv\Theta(-1)$ (dependent on $\eta$),
obtained easily by calculating the limit $\phi\to+\infty$ in
(\ref{form:teta fi eta}):
\begin{equation}\label{form:Teta min}
\Theta_\mathrm{min}=3f^{-1}(\eta) 
\left[1-\frac{(\cosh\eta-1)^2}{\sinh\eta\, (\sinh\eta-\eta)}\right] .
\end{equation}
For $\eta\to0$, equivalent to $\Omega\to1$ (the
Einstein--de Sitter model of the universe), we get the value of
$\Theta_\mathrm{min}=-1.5$, which can also be calculated
otherwise.\footnote{In the E--dS model we have
  $H = 2\slash(3t)$ and $f(\Omega)\equiv1$; adopting the
  empty world model (Milne model) for the perturbation, we get
  $H_\mathrm{p}=t^{-1}$ and further on
  $\Theta=\theta=3(1-H_\mathrm{p} H^{-1})=-1.5$.} The
opposite limit of $\Omega\to0$ ($\eta\to+\infty$)
leads to $\Theta_\mathrm{min}\to-1$; this can be equally deduced from (\ref{eq:linear theory}). If we adopt the currently accepted
value of $\Omega_0\simeq0.25$ ($\eta_0\simeq2.63$), we obtain
$\Theta(-1)\simeq-1.43$. Thus, the \citet{B92} approximation (\ref{form:Bernard Teta}),
which gives $\Theta(-1)=-1.5$ independently of $\Omega$, has a
relative error of approx. 5\% in this limit for such $\Omega_0$.

We would now like to find an (approximate) relation $\Theta$--$\delta$ for
the whole range $\delta \in [-1,0]$. \citet{B92} derived his formula
expanding the relation around $\delta = 0$. We adopt a different
approach: we expand the relation around $\delta = -1$. (That is, at a
first step we assume $0\le \delta + 1 \ll 1$). Then, for arbitrary
$\eta$, the perturbation parameter $\phi \gg 1$. From
Equation~(\ref{form:delta fi eta}) we obtain $\cosh\phi_1 =
g(\eta)/(1+\delta)$, where

\begin{equation}
g(\eta) = \frac{(\cosh\eta - 1)^3}{(\sinh\eta - \eta)^2} \,,
\label{eq:g_eta}
\end{equation}
and further on
\begin{equation}
\cosh\phi_2 = \cosh\phi_1 + 3 - 2\phi_1 \,.
\label{eq:phi2_eta_arbitrary}
\end{equation}
Using Equation~(\ref{eq:phi2_eta_arbitrary}) in 
Equation~(\ref{form:teta open}) we get 

\begin{equation}
\Theta_2 = \Theta_{\rm min} + f^{-1}(\eta) \frac{3\, (\sinh\eta - \eta)\, (\phi_1 -
  2)}{\sinh\eta\, (\cosh\eta - 1)}\, (1 + \delta) \,,
\label{eq:Theta2_eta_arbitrary}
\end{equation}
where
\begin{equation}
\phi_1 = \ln\left[2g(\eta)\right] - \ln(1+\delta) \,.
\label{eq:phi1_eta_arbitrary}
\end{equation}
Equation (\ref{eq:Theta2_eta_arbitrary}) satisfies explicitly the
highly non-linear limit $\Theta(-1) = \Theta_{\rm min}$. Also, in the
limit $\eta \gg 1$, this Equation reduces to asymptotic
Equation~(\ref{form:tetafal2}), as expected.

The range of applicability of formula~(\ref{eq:Theta2_eta_arbitrary})
is very limited: it starts to deviate from the exact relation for
$\delta$ about $-0.9$. We would like to introduce such a modification so as to satisfy
also the linear-theory limit: for $|\delta|\ll1$,
$\Theta=\delta$. Therefore, we adopt the following three boundary
conditions:

\begin{enumerate}
\item $\Theta(-1) = \Theta_{\rm min}$,
\item $\Theta(0)=0$,
\item \mbox{$\left(\de\Theta\slash\de\delta\right)|_{\delta=0}=1$}.
\end{enumerate}
Inspired by Equation~(\ref{eq:Theta2_eta_arbitrary}), we write

\begin{equation}
\Theta = \Theta_{\rm min} + a_1(\eta) (1 + \delta) + a_2(\eta) 
(1 + \delta) \ln(1 + \delta) \,,
\label{eq:Theta2_general}
\end{equation}
where $a_1$ and $a_2$ are arbitrary functions of $\eta$. Imposing
the three boundary conditions A.--C. on the above formula we find

\begin{equation}
\Theta=\delta + (1+\Theta_\mathrm{min})\, N(\delta)\,, 
\label{eq:Theta2_ansatz}
\end{equation}
where $\Theta_\mathrm{min}$ as a function of $\Omega$ [cf. (\ref{eq:Theta_min})] is
\begin{displaymath}
\Theta_\mathrm{min}(\Omega)=3\, f^{-1}(\Omega)\, \times
\end{displaymath}
\begin{displaymath}
\times\left\{
1-\frac{2\sqrt{(1-\Omega)^3}}{2\sqrt{1-\Omega}-\Omega\ln[2\Omega^{-1} 
(1+\sqrt{1-\Omega})-1]}\right\} \simeq
\end{displaymath}
\begin{equation}
\simeq \, -1 - 0.5 \Omega^{0.12 - 0.06\, \Omega}
\label{eq:Theta_min}
\end{equation}
[here $f(\Omega)$ is given by (\ref{form:f of Omega})] and
$N(\delta)$ has the form of (\ref{form:N of delta}). Indeed,
formula~(\ref{eq:Theta2_ansatz}) meets all the three boundary
conditions: the last two are fulfilled since for small $\delta$,
$N(\delta) = \delta^2/2 + \ldots$, and the first one because $N(-1) =
1$.  This simple approximation is robust for $\delta$ close to
$-1$ and around $0$; for intermediate values of $\delta$ it slightly
underestimates the exact value of $\Theta$ (with a maximal relative
error of 2\% for $\Omega\simeq0.25$).

\begin{figure} 
\begin{center}
\includegraphics[width=0.5\textwidth]{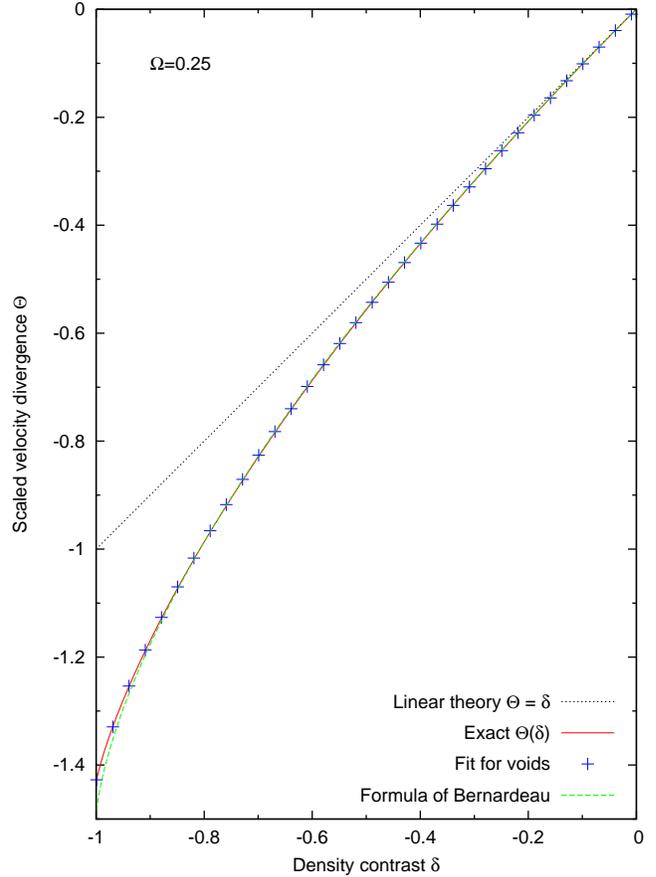} 
\end{center}
\caption{\label{fig:fit voids} Relation between density contrast and
  scaled velocity divergence for spherical voids (underdensities):
  linear theory (dotted line), exact relation (solid line),
  \citet{B92} approximation (dashed line) and fit given by
  Equation~(\ref{eq:Theta3_ansatz}) (crosses). The density parameter
  of the background is $\Omega=0.25$.}
\end{figure}

Formula~(\ref{eq:Theta2_ansatz}) is probably already sufficiently
accurate for practical applications. Still, it is of course possible
to improve it. In order to do this, we expand the exact
$\Theta$--$\delta$ relation around $\delta = -1$ up to third-order in
the perturbation parameter $\phi$. The result is the following series:
\begin{eqnarray}
\Theta_3 = a_0 + a_1 (1+\delta) + a_2 (1+\delta) \ln(1+\delta) +
a_3(1+\delta)^2 && \nonumber \\
+\, a_4 (1+\delta)^2 \ln(1+\delta) + a_5 (1+\delta)^2 \ln^2(1+\delta) 
, &&
\label{eq:Theta3_general}
\end{eqnarray}
where $a_i$ are some functions of $\eta$ (see
Appendix~\ref{app:Theta3}). From the six terms above we construct
their linear combinations which fulfill the constraints A.--C. This, together with the condition of simplicity, leads us to postulate
\begin{eqnarray}
\Theta=\delta + \left[1+\Theta_\mathrm{min}(\Omega)\right]N(\delta) +
\alpha_1(\Omega)\, \delta\, (1+\delta) \ln(1+\delta) \,+\, && \nonumber \\
\alpha_2(\Omega) (1+\delta)^2 \ln^2(1+\delta)\,, \hphantom{x} && 
\label{eq:Theta3_ansatz}
\end{eqnarray}
Fitting this formula to the exact relation gives $\alpha_1 = 0.12$ and
\mbox{$\alpha_2 = -0.09$} for $\Omega = 0.25$. The fit is very accurate: it
has a maximal error smaller than $0.2$\%. In general, both $\alpha_1$ and $\alpha_2$ depend weakly on $\Omega$: we have $\alpha_1\simeq0.19\,\Omega^{0.35}$ and $\alpha_2\simeq-0.15\,\Omega^{0.35}$ for \mbox{$0.1\leq\Omega\leq0.9$.} Figure \ref{fig:fit voids}
presents a comparison of the exact relation for $\Theta(\delta)$,
calculated for voids from (\ref{form:teta open}), with the fit
(\ref{eq:Theta3_ansatz}) and the \citet{B92} approximation (\ref{form:Bernard Teta}). As we can see, our fit lies accurately on the
approximated curve and the \citet{B92} formula slightly underestimates
exact values of $\Theta$ for $\delta$ close to $-1$.  However, it
should be admitted that the latter is considerably simpler than
ours.

%%%%%%%%%%%%%%%%%%%%%%%%%%%%%%%%%%%%%%%%%%%%%%%%%%%%%%%%%%%%%%%%%%%%%%%%%%%%%%%%%%%%%%%%%
\section{Overdensities}
\label{sec:overd}

An \textit{overdensity} is any perturbation for which $\delta>0$. As
already mentioned, these can be of two types, depending on the initial
density contrast: `open' or `closed'. For a specific value of
$\Omega$ (or, equally, $\eta$), the boundary value of the density
contrast, maximal for the first type and minimal for the second, is
given by (\ref{form:delta lim}). For the currently accepted
value of $\Omega_0\simeq0.25$ we have
$\delta_\mathrm{lim}\simeq1.6$: such overdense but open perturbations
($\delta<\delta_\mathrm{lim}$) fall within the weakly non-linear
regime.

In order to find an approximation for $\Theta(\delta)$ for overdense
spherical regions, we will use a similar procedure as we did for
voids, examining the highly non-linear regime ($\delta \gg 1$). Owing
to the considerations above, it is sufficient to focus on closed
perturbations; the formula for $\Theta$ is then of the form
(\ref{form:teta closed}), with the `+' sign. Highly non-linear infall
means that the overdensity collapses to a point: the conformal time of
the perturbation $\phi\to2\pi$. This is in general not
physical, as in practice for $\phi\la 2\pi$ virialisation would occur
and prevent further collapse. However, as in the case of voids,
examination of this regime leads to interesting formulae.

First of all, we can directly put $\phi=2\pi$ into (\ref{form:teta
closed}), getting the `1-st order approximation':
\begin{equation}\label{form:Teta 1}
\Theta_1=3f^{-1}(\Omega)\left[1+\sqrt{\Omega(1+\delta)}\right]\,.
\end{equation}
 We can see that the \citet{B92} formula (\ref{form:Bernard Teta}),
 which was not intended to work in this regime, indeed will not work:
 already the slope of the curve is incorrect (2/3 instead of 1/2). For
 realistic values of $\Omega$,
 $f^{-1}(\Omega)\sqrt\Omega\simeq\Omega^{-0.1}$. Using this
 approximate equality and neglecting the constant term in
 Equation~(\ref{form:Teta 1}) yields the `0-th order approximation',
 $\Theta_0 = 3\, \Omega^{-0.1} \sqrt{1+\delta}$. The same relation can
 be deduced from dynamical considerations (namely, from energy
 conservation in the highly non-linear infall). \citet{NuCo} also found such a form of the weak $\Omega$-dependence
 $\left(\Omega^{-0.1}\right)$ of the peculiar velocity in virialised
 regions. This is not surprising, since both in our and their case,
 $\delta \gg 1$ and $\Theta \ll \delta$.

\begin{figure} 
\begin{center}
\includegraphics[angle=270, width=0.5\textwidth]{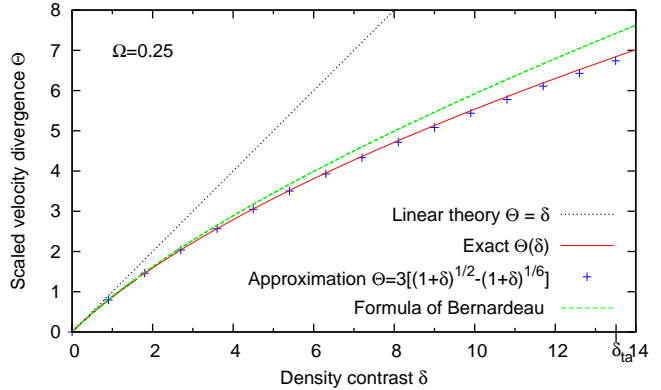} 
\end{center}
\caption{\label{fig:overdense to} A comparison of proposed
  approximations for $\Theta(\delta)$ with the \mbox{exact} relation for
  spherical overdensities in the mildly non-linear regime (up to the
  turn-around). Solid line shows the exact relation, the \citet{B92}
  approximation~(\ref{form:Bernard Teta}) is illustrated by dashed
  line and crosses present the fit given by (\ref{form:fit pot});
  dotted line is the linear theory relation. The density parameter of
  the background equals to $\Omega_0=0.25$; the density contrast of
  the turn-around is then
  $\delta_\mathrm{ta}\simeq13.5$.}
\end{figure}

Expanding the relation (\ref{form:teta closed}) around $\phi=2\pi$
($0\le 2\pi - \phi \ll 1$) to higher order, we obtain the following
series:
\begin{eqnarray} \label{form:Teta ser kolaps}
\Theta=3f^{-1}(\Omega)\big[1+\sqrt\Omega(1+\delta)^{1\slash2} + 
a_{1\slash6}\sqrt\Omega(1+\delta)^{1\slash6} + \nonumber \\ 
+a_{-1\slash6}\sqrt\Omega(1+\delta)^{-1\slash6}+\cdots\big] \,,
\end{eqnarray} 
where $a_\mathrm{i}$ are functions of $\Omega$ only. In order to
obtain a fit that would both converge to (\ref{form:Teta 1}) in the
highly non-linear regime of $\delta\gg1$ and have proper behaviour in
the vicinity of $\delta=0$ (conditions B. and C. from Section \ref{sec:voids}), we proceed similarly as we did for
$\delta<0$. First, already here we neglect the fourth (and all
next) component of the series. Then we modify the expansion
(\ref{form:Teta ser kolaps}) by introducing two parameters
$\mathcal{A}$, $\mathcal{B}$ and an integer $n$:

\begin{equation}\label{form:fit A,B,n}
\Theta=3f^{-1}\left[\mathcal{A}+\sqrt{\Omega}\sqrt{1+\delta} - 
\mathcal{B}\sqrt{\Omega}(1+\delta)^{1\slash n}\right] .
\end{equation}
Imposing the conditions $\Theta(0)=0$ and
$\left(\de\Theta\slash\de\delta\right)|_{\delta=0}=1$, we obtain:
\begin{equation}
\mathcal{A}=\left(\frac{n}{2}-1\right)\sqrt\Omega - 
\frac{n}{3}f(\Omega)\,,\qquad
\mathcal{B}=\frac{n}{2}-\frac{n}{3\sqrt\Omega}f(\Omega)\,.  
\end{equation} 
In particular, for $n = 6$ [cf. (\ref{form:Teta ser kolaps})] we have
\begin{equation}
\mathcal{A}(\Omega,n=6) \simeq 2 \Omega^{0.5}\! \left(1 - \Omega^{0.1}\right),
\end{equation}
and
\begin{equation}
\mathcal{B}(\Omega,n=6) \simeq 3 - 2 \Omega^{0.1}
\label{form:A,B,6}
\end{equation}
(remembering that $f\simeq\Omega^{0.6}$). Inserting the above
expressions for $\mathcal{A}$ and $\mathcal{B}$ into
Equation~(\ref{form:fit A,B,n}) with $n=6$ and neglecting the weak
\mbox{$\Omega$-dependence} [since $f^{-1}(\Omega)\sqrt{\Omega}\simeq\Omega^{-0.1}$],
we obtain the
following result, which can be treated as a generalisation of the
\citet{B92} formula:
\begin{equation}\label{form:fit pot}
\Theta=3\left[(1+\delta)^{1/2}-(1+\delta)^{1/6}\right]\,.
\end{equation}
An interesting feature of this fit is that it has the same
second-order Taylor expansion around $\delta=0$ as the approximation
given by \citet{B92}:
\begin{equation}
\Theta=\delta-\frac{1}{6}\delta^2+\cdots\,.  
\end{equation}
This means that in the weakly non-linear regime these two approximations give
similar results. However, for mildly non-linear values of $\delta$
(from $\delta_\mathrm{lim}$ up to the turn-around\footnote{The
  \textit{turn-around} of a closed perturbation is the moment when it
  stops expanding, i.e. $\dot{a}_\mathrm{p}=0$. In the spherical model
  as discussed here it occurs for $\phi=\pi$; the density contrast for
  the turnaround spans from $\delta_\mathrm{ta} = 9\pi^2/16 -1 \simeq
  4.6$ for $\Omega = 1$ to $\delta_\mathrm{ta}\simeq 30$ for
  $\Omega=0.1$. If $\Omega=0.25$, then $\delta_\mathrm{ta}\simeq 13.5$.}) our approximation works generally better than the formula
of \citet{B92}. The maximal error of our fit for such an interval of
density contrasts is about 1.5\%. Figure~\ref{fig:overdense to}
shows the discussed approximations for the weakly and mildly
non-linear regime.

\begin{figure} 
\begin{center}
\includegraphics[angle=270, width=0.5\textwidth]{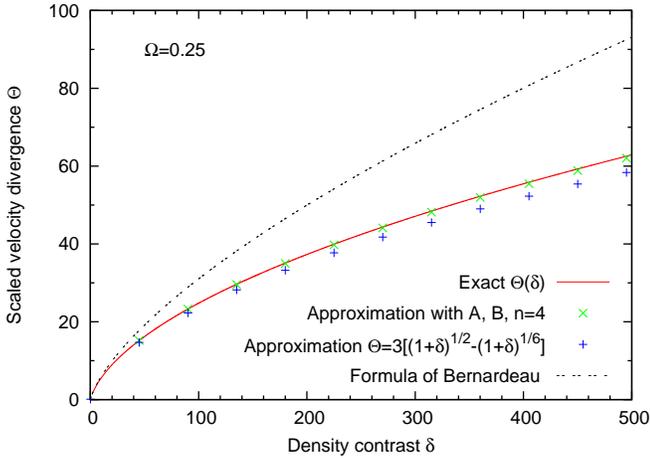} 
\end{center}
\caption{\label{fig:overdense hnl} An illustration of the behaviour of
the function $\Theta(\delta)$ for spherical overdensities in the
highly non-linear regime, i.e. up to the virialisation. The solid curve
is the exact relation, the dashed line shows the \citet{B92}
approximation and crosses present two approximations described in the
text: plus marks represent the formula (\ref{form:fit pot}) and
multiplication marks show the fit (\ref{form:fit A,B,n}) with
$n=4$.}
\end{figure}

For higher values of $\delta$, neither the \citet{B92} approximation,
nor the fit (\ref{form:fit pot}) are adequate. In case of the first
one this is mainly due to a wrong slope of the curve; in case of the
second -- due to the negligence of the dependence on $\Omega$. For
that reason in the regime of very big $\delta$ we prefer to use the
fit (\ref{form:fit A,B,n}), of a more general form. The approximation (\ref{form:fit pot})
suggests that the best choice of $n$ is $n=6$; however, it turns out
that in practice, for highly non-linear density contrasts (greater
than the value for the turn-around), approximation~(\ref{form:fit A,B,n}) with $n = 4$ works slightly
better than with $n = 6$ (with the weak $\Omega$-dependence included
in both cases). In Figure~\ref{fig:overdense hnl} we show the
behaviour of the function $\Theta(\delta)$ in the highly non-linear
regime. For comparison, we plot the formula of \citet{B92}, the simple approximation~(\ref{form:fit pot}) and
the approximation~(\ref{form:fit A,B,n}) with $n = 4$.

Our results for the highly non-linear regime are rather of academic
value, since, as stated earlier, highly non-linear infall is
considerably modified by the effects of virialisation. To account for
them (and for deviations from spherical symmetry), \citet{ShMo} constructed an {\em improved\/} (extended) semi-analytical
spherical collapse model. For $\delta$ up to about
$\delta_\mathrm{ta}$ (which equals to $\sim4.6$, as the background assumed in the discussed paper is of the Einstein--de Sitter type) their model coincides with the standard
spherical model (studied here), while for larger density contrasts it
deviates from the latter and under some additional assumptions matches
well the results of N-body simulations presented by \citet{Hami}. Indeed, formula~(20) of \citet{ShMo}, for $T = \tau$
(the limit of the standard model), reduces to our Equation~(\ref{form:teta closed}) (their $h_\mathrm{SC} = \theta/3$). The authors argue that for background universes with dark energy their formula is
valid only for $\delta \ga 100$. They claim that for smaller values
of $\delta$, their results are not accurate. We disagree with these
statements. As already stated, the weak $\Omega$ {\em and\/} $\Lambda$
dependence of the scaled velocity--density relation has been shown on
the level of the equations of motion \citep{NuCo}, so
independently of the level of non-linearity. Since for small redshifts
dark energy behaves similarly to the cosmological constant
(e.g. \citealt{Riess}), and since only for such redshifts the weak
dependence of equations of motion on cosmological parameters
starts to play any role (because earlier we had $\Omega \la 1$; \citealt{NuCo}), the velocity--density relations for cosmological models with
and without dark energy must be similar.

%%%%%%%%%%%%%%%%%%%%%%%%%%%%%%%%%%%%%%%%%%%%%%%%%%%%%%%%%%%%%%%%%%%%%%%%%%%%%%%%%%%%%%%%%%%%
\section{Comparisons with fits to numerical simulations}
\label{sec:comp}
\citet{KaCPeR} studied the mildly non-linear
velocity--density relation using the Cosmological Pressureless
Parabolic Advection (CPPA) hydrodynamical code. They found that the
mean relation between the scaled velocity divergence and the density
contrast can be very well described by the so-called `$\gamma$-formula',

\begin{equation}\label{eq:Theta_gamma}
\Theta = \gamma \left[(1+\delta)^{1/\gamma} - 1\right] + \eps \,,
\end{equation}
with $\gamma \simeq 1.9$. This formula is a modification of the \citet{B92}
formula with $\gamma$ instead of $3/2$. The offset $\eps > 0$ is
introduced to account for an effect of a finite variance of the
density field: the value of $\eps$ is such that the global mean of
$\Theta$ is zero, as required. (Another effect of a finite variance is
to modify the degree of non-linearity of the relation.) Without the offset, the above formula yields $\Theta(-1)
= -\gamma = -1.9$ for $\gamma = 1.9$, in significant difference with
the value $-1.5$, obtained neglecting the weak \mbox{$\Omega$-dependence} of the
exact limit, Eq.~(\ref{form:Teta min}). However, for Gaussian
smoothing scales of a few Mpc, employed in \citet{KaCPeR}, the offset shifts
the value of $\Theta(-1)$ much closer to $-1.5$.

\citet{B99} analysed the
velocity--density relation using N-body si\-mu\-la\-tions performed for
various background cosmologies. They noticed a weak dependence of the
relation on $\Omega$ and $\Lambda$. \citet{B99} invented a somewhat more
elaborate fit to the extracted mean relation, presented in the form of
density in terms of velocity divergence,

\begin{equation}
\delta = \beta \left(1 + \Theta/\gamma \right)^\gamma - 1 \,.
\label{eq:delta_B99}
\end{equation}
Here, $\beta$, slightly smaller than unity, plays a role of the offset
$\eps$ in Equation~(\ref{eq:Theta_gamma}): it assures that the global
mean of $\delta$ is zero, as required. In
Equation~(\ref{eq:delta_B99}) $\gamma$ is not a constant, but is
approximated as a following function of $\Theta$:
\begin{equation}
\gamma = \frac{3}{2} + 0.3\, \Omega^{0.6} \left(\Theta + \frac{3}{2}
\right) \,.
\label{eq:gamma}
\end{equation}
The above equation quantifies the fact that for larger values of
velocity divergence, the observed relation becomes more
non-linear. Indeed, $\gamma$ grows with growing $\Theta$ (we recall
that $\gamma = 1$ corresponds to the linear theory). Moreover, for
$\Theta = - 3/2$, we have $\gamma = 3/2$, so then $\delta = -1$, as it was
intended. [Note a typo in eq.~(20) of \citet{B99}: instead of $\tilde\theta$ (in
  our notation, $\Theta$), there should be $\theta$.]

How do these findings, based on fully non-linear simulations, relate to
our results? In overdensities, our Formula~(\ref{form:fit pot})
follows closer to the exact relation in the SCM than the formula of
\citet{B92}. Moreover, our approximation is a formula with increasing effective
index $\gamma_{\rm eff}$. Its second order expansion is the same as
that of \citet{B92}, so for small $\delta$, $\gamma_{\rm eff} = 3/2$. For
large density contrasts, the second term in Equation~(\ref{form:fit
  pot}) becomes negligible, so asymptotically $\gamma_{\rm eff} = 2$
(for $\delta \gg 1$). Therefore, qualitatively our formula is
consistent with the fit of \citet{B99}, in a sense that $\gamma$, as a
function of $\Theta$ or $\delta$, is growing. It is also consistent
with the fit of \citet{KaCPeR}, in a sense that the average $\gamma$ is slightly
larger than $3/2$. Clearly, our formula is a better fit to the results
of numerical simulations than the formula of \citet{B92}.

Of course, quantitatively there are discrepancies. First of all, it is
strictly impossible to satisfy simultaneously the features of both
fits: $\gamma$ is {\em either\/} constant {\em or\/} increasing. This
discrepancy between the results of the two groups is not necessarily a
sign of a major flaw in any of their analyses. The two groups used
different codes: N-body versus hydro. The first one follows accurately
non-linear evolution, but provides naturally a mass-, not volume-,
weighted velocity field, while the latter is needed. CPPA, as any
hydrodynamical code, provides naturally a volume-weighted velocity
field, but follows the non-linear evolution after shell crossings only
approximately. Moreover, the density power spectra used in both
simulations were different. Also, fit~(\ref{eq:delta_B99}) of \citet{B99} was
found for top-hat smoothed fields, while fit~(\ref{eq:Theta_gamma}) of
\citet{KaCPeR} was elaborated for fields smoothed with a Gaussian filter (more
appropriate for velocity--density comparisons). The effects of
smoothing, though small, are different for these two filters (see
e.g.\ Table~1 of \citealt{KaCPeR}). Finally, an inverse of the forward relation
(density in terms of velocity divergence) does not strictly describe
the mean inverse relation, due to scatter.

Which results better reflect real non-linear dynamics of cosmic random
density and velocity fields? Instead of betting, it would be probably
best to repeat the analysis using an output from high-resolution
N-body simulations with a $\Lambda$CDM power spectrum, employing --
instead of a Voronoi tessellation \citep{BvdW}
-- a much simpler algorithm of extracting volume-weighted velocity
field of \citet*{CCT}. Voronoi tessellations
are complicated and very CPU-consuming, so they can be applied only to
a limited number of points, while the method of \citet{CCT} can be (and actually has been) applied to all simulation points
($512^3$ in their work). If the actual relation is not more non-linear
than in the highly non-linear regime of the SCM ($\gamma = 2$), then we
can use Formula~(\ref{form:fit A,B,n}), with neglected weak
$\Omega$-dependence and $n$ treated as a free parameter. Let us write
it explicitly:
\begin{equation}\label{eq:Theta_fit_n}
\Theta^{(2,n)} = 3 \left[(1+\delta)^{1/2} - \frac{n}{6}
  (1+\delta)^{1/n}\right] + \eps \,,
\end{equation}
where $\eps = n/2 - 3$. [For $n = 6$, it reduces to Formula~(\ref{form:fit pot}).] For example, if the best-fit value of $\gamma$
is found to be close to $1.9$ and fairly constant, then $n = 2.3$ 
would provide an excellent fit. Instead, significant `run' of the index
$\gamma$ would probably demand $n > 6$. 

If, on the other hand, the results of \citet{B99} are found to be accurate, then for
$\Omega = 0.25$ Equation~(\ref{eq:gamma}) yields $\gamma = 2$ already
for $\Theta \simeq 2.3$, and even more for higher $\Theta$. In this case,
in order to describe the relation up to the turn-around, one should modify
also the exponent of the leading term in
Formula~(\ref{eq:Theta_fit_n}) ($1/m$ instead of $1/2$, with $m\ga
2$).\footnote{This modification would create a coefficient of the
  leading term equal to $m/2$ and modify the offset to $\eps = n/2 -
  (3m/2)$.} It is a matter of choice if to fit one `running' exponent
($\gamma$) or two constant ($m$ and $n$). In any case, it is better to
use an additive offset $\eps$ instead of the factor $\beta$, appearing
in Equation~(\ref{eq:delta_B99}): in applications to
velocity--velocity comparisons the value of $\eps$ is not relevant at
all. The mildly non-linear velocity field is vorticity-free to good
accuracy, so the predicted velocity field (from the density field) is
\begin{equation}
 \bfv(\bfr)=
   \frac{H f(\Omega)}{4\pi}\!\! \int\! {\rm d}^3\bfr' \,\,
   \frac{\Theta[\delta(\bfr')]\, (\bfr'-\bfr)}{|\bfr'-\bfr|^3}
 \,,
 \label{eq:theta_int}
\end{equation}
and the contribution of the offset to velocity averages out to zero. 

An advantage of the $\gamma$-formula over
Formula~(\ref{eq:Theta_fit_n}) or its modification is that it works
also for voids. For underdensities, the formula of \citet{B92} is a very good
description of the exact relation in the SCM, except for the very tail
$\delta \simeq -1$ (where the weak $\Omega$-dependence becomes important). Results of numerical simulations show very limited need to
modify the formula of \citet{B92} for voids -- the discrepancies appear at
larger density contrasts. As stated earlier, \citet{B92} predicted this
fact. Our formulae for voids give results very similar to that of \citet{B92},
but describe better the regime $\delta \simeq -1$. This regime is
important for predicting expansion velocities of almost completely
empty voids (e.g., see \citealt{Tully}). Therefore, using
approximation~(\ref{eq:Theta2_ansatz}) for underdensities, we propose the
following combined formula:
\begin{equation}
\Theta = \left\{ \begin{array}{lr} \Theta^{(m,n)} \,, & \delta > 0 , \\ 
\delta - 0.5\, \Omega^{0.12 - 0.06\, \Omega} N(\delta) + \eps \,,  
& -1 \le \delta \le 0 \,.
\end{array} \right. 
\label{eq:Theta_comb}
\end{equation}
Here, $\Theta^{(m,n)}$ is the `$m$-modification' of
Formula~(\ref{eq:Theta_fit_n}), $N(\delta)$ is given by
Equation~(\ref{form:N of delta}) and $\eps$, treated as a free
parameter, is the same in both cases. Alternatively, as the relation
for voids, one could use more complicated and extremely accurate
Equation~(\ref{eq:Theta3_ansatz}). To sum up, we admit that it is
disputable if to fit results of numerical simulations using our
formula~(\ref{eq:Theta_comb}), or $\gamma$-formula. What remains
indisputable is that for overdensities, our standard formula (with $m
= 2$ and $n = 6$) is a better starting fit than the standard formula
of \citet{B92} (with $\gamma = 3/2$).

%%%%%%%%%%%%%%%%%%%%%%%%%%%%%%%%%%%%%%%%%%%%%%%%%%%%%%%%%%%%%%%%%%%%%%%%%%%%%%%%%%%%%%
\section{Summary and conclusions}
\label{sec:concl}
The main motivation of this paper was to rederive the formula of \citet{B92}
in a simple way, using the spherical collapse model (SCM), and to
extend it to larger density contrasts, where it is no longer
valid. The undertaken project abounded in surprises:
\renewcommand{\labelenumi}{\roman{enumi}.}
\begin{enumerate}
  \item Contrary to the claim of \citet{B99}, the formula of \citet{B92} is not exact in the
    limit of an empty universe. On the contrary, it completely fails
    in this regime: the exact relation in the SCM is then $f^{-1}
    \nabla \cdot \mathbf{v} = \delta$, for an arbitrary $\delta$. In
    fact, this is a general result of dynamics in a low-density
    universe.
  \item Although the formula of \citet{B92} fails for $\Omega \to 0$, where it
    was expected to work best, for realistic values of $\Omega$ (say,
    $\Omega > 0.1$), it describes very well the SCM velocity--density
    relation in voids. It also works for overdensities up to
    $\delta \sim 2$ -- $3$.
\end{enumerate}

\ind The velocity--density relation in the SCM is given in a parametric
form. Our goal here was to eliminate this parameter (at least
approximately) and to provide the relation analytically. We
aimed at describing the relation in the whole range $\rho \in (0,
\infty)$ (realistically, up to $\rho_{\rm vir}$). Therefore, instead
of expanding it around $\rho = \rho_{\rm b}$, we adopted an entirely
different approach. Namely, we derived asymptotes of the relation in
the highly non-linear regime: $\rho/\rho_{\rm b} \gg 1$ ($\delta \gg
1$) for overdensities and $\rho_{\rm b}/\rho \gg 1$ \mbox{($0 \le 1 + \delta
\ll 1$)} for voids. (Although we also `expanded' around them, in a
sense that we also calculated next-to leading-order terms.) These two
asymptotes turned out to be qualitatively different. Inspired by their
functional forms, we invented semi-phenomenological fits to the exact
relation (separately for overdensities and voids), fulfilling the
linear theory condition $f^{-1} \nabla \cdot \mathbf{v} = \delta$.\\
\ind For overdensities, our main result is Formula~(\ref{form:fit pot}). It
describes well the exact relation in the SCM up to the turn-around (for
$\Omega = 0.25$, $\delta_{\rm ta} = 13.5$). As already stated, the
formula of \citet{B92} starts to deviate from the exact relation for $\delta
\sim 3$. We have also fitted the regime $\delta \in (\delta_{\rm
  ta},\delta_{\rm vir})$, though virialisation and departures from
spherical symmetry make practical applicability of the SCM in this
regime very limited.\\
\ind In case of voids, the most important results of this paper are Formulae~(\ref{eq:Theta2_ansatz})
and~(\ref{eq:Theta3_ansatz}), with $\Theta_\mathrm{min}$ given by
Equation~(\ref{eq:Theta_min}). Compared with the SCM, simple
Formula~(\ref{eq:Theta2_ansatz}) has a maximal error of about $2$\%
and is probably sufficient for practical applications. The formula of
\citet{B92} is an even better approximation, except for the limit $\delta \to
-1$, where for $\Omega = 0.25$ it has approximately $5$\% relative
error. Our more complicated formula~(\ref{eq:Theta3_ansatz}) is
extremely accurate in the whole range $\delta \le 0$: its maximal
error is about $0.2$\%.\\
\ind An ultimate goal of studies such as the present one is to find the
relation valid for realistic random cosmic velocity and density
fields. Unlike the work of \citet{B92}, our calculations were greatly
simplified by the strong assumption of spherical symmetry. There is
therefore no guarantee that better agreement with the SCM implies
better agreement with the real relation. In order to check this issue
we compared our formulae to fits to results of cosmological numerical
simulations, that are present in the literature. We have found that
in voids, our formulae, as well as the formula of \citet{B92}, describe well the
real relation. This is partly a consequence of the fact that voids are
more spherical than overdensities. In overdensities, both our formula
and that of \citet{B92} require modification, but ours less. This discrepancy
is not a failure of the latter of the two, since it has never
been intended to work for $\delta \ga 2$. Our formula~(\ref{form:fit
  pot}), having the same second-order expansion as the formula of \citet{B92},
can be regarded as its extension into the mildly non-linear regime (for
$\delta$ up to the turn-around). We have also discussed how to (slightly)
modify our formula to better fit numerical simulations.\\
\ind In Section~\ref{sec:intr} we have enlisted arguments for weak
dependence of the velocity--density relation on the cosmological
parameters. Therefore, in the present analysis we set $\Lambda =
0$. To study the limit $\Omega \to 0$ we have retained
$\Omega$-dependence of the equations of the SCM. Analysing these
equations we have confirmed that for realistic values of $\Omega$, the
$\Omega$-dependence of the relation is indeed very weak. In final
formulae it has been therefore neglected, except for
Formula~(\ref{eq:Theta_min}) for $\Theta_\mathrm{min}$. The difference
between $\Theta_\mathrm{min}$ for $\Omega = 1$ and $\Omega = 0.25$ is
about $5$\%. In fact, if we want to have better accuracy, there is no
guarantee that $\Lambda$ does not contribute at a comparable level. It
is then worth to repeat the analysis with $\Lambda = 1 - \Omega$. We
plan to undertake such a study in the future.

\section*{Acknowledgments}
This work was partially supported by the Polish Ministry of Science and
Higher Education under grant N N203 0253 33, allocated for the period
2007--2010.

%%%%%%%%%%%%%%%%%%%%%%%%%%%%%%%%%%%%%%%%%%%%%%%%%%%%%%%%%%%%%%%%%%%%%%%%%%%%%%%%%%%%%%%%%%%%%%%%%%%%%%%%%%%
%\newpage

\appendix

\section{The velocity--density relation in an empty universe}
\label{app:empty}

The general equation of motion for the cosmic pressureless fluid in
comoving coordinates is
\begin{equation}
\frac{\partial \bfv}{\partial t} + \frac{1}{a} \left(\bfv \cdot
\nabla\right) \bfv + \frac{\dot{a}}{a} \bfv = \bfg \,,
\label{eq:motion}
\end{equation}
where $\bfg$ is the peculiar gravitational acceleration,
\begin{equation}
 \bfg(\bfx,t) = G \rho_{\rm b} a \!\! \int\! {\rm d}^3\bfx' \,\,
   \frac{\delta(\bfx',t)\, (\bfx'-\bfx)}{|\bfx'-\bfx|^3}
 \label{eq:grav}
\end{equation}
(e.g.\ Peebles 1980). For $|\delta| \ll 1$ we can neglect the
non-linear term on the LHS of Equation~(\ref{eq:motion}). Let us choose
some instant of time, $t_i$, of the evolution of an open universe when
already $\Omega \ll 1$. For such $\Omega$ perturbations stop growing,
so for $t > t_i$, \mbox{$\bfg(\bfx,t) = \bfg_i(\bfx)/a^2$.} Our
Equation~(\ref{eq:motion}) simplifies then to
\begin{equation}
\frac{\partial}{\partial t}[a\, \bfv(\bfx,t)] = \frac{\bfg_i(\bfx)}{a} .
\label{eq:motion_empty}
\end{equation}
The solution is 
\begin{equation}
\bfv(\bfx,t) = H_0^{-1} (\eta - \eta_i) \frac{\bfg_i(\bfx)}{a(t)} + 
\frac{\bfv_i(\bfx)}{a(t)} + \frac{\bmath{F}(\bfx)}{a(t)},
\label{eq:solut}
\end{equation}
where the conformal time $\eta$ is in general defined in
Equation~(\ref{form:conft}). The last term in Equation~(\ref{eq:solut})
is the homogeneous part. Here we do not assume {\em a priori\/}
irrotationality of the velocity field, so we retain this term. (Though
it does not contribute to the velocity divergence, because $\nabla
\cdot \bmath{F} = 0$.) The limit $\Omega \to 0$ corresponds to $\eta
\to \infty$. Therefore, in the above equation we can neglect the terms
$\bfv_i/a$ and $\bmath{F}/a$, as well as $\eta_i$. This yields
\begin{equation}
\bfv = H_0^{-1} \eta\, a\, \bfg \,.
\label{eq:solut_limit}
\end{equation}
From Equation~(\ref{eq:grav}) we have
\begin{equation}
\nabla \cdot \bfg = - 4\pi G \rho_{\rm b} a \delta = - \frac{3}{2} H^2
\Omega a \delta \,.
\label{eq:div_g}
\end{equation}
This yields in~(\ref{eq:solut_limit}) 
\begin{equation}
\nabla \cdot \bfv = - \frac{3}{2} H_0^{-1} (H a)^2 \Omega \eta\, \delta \,.
\label{eq:div_v}
\end{equation}
In an (almost) empty universe $H(t) = t^{-1}$ and $a(t) = t/t_0$,
hence $H a = H_0$. Also, the general relation~(\ref{form:Omega of
  eta}) between $\Omega$ and the conformal time simplifies then to
$\Omega = 4 {\rm e}^{-\eta}$. Substituting this in
Equation~(\ref{eq:div_v}) we obtain $\nabla \cdot \bfv = - H_0 6 \eta
{\rm e}^{-\eta} \delta$. Comparing this equation with Equation~(\ref{form:f eta big}), we
identify the factor $6 \eta {\rm e}^{-\eta}$ as the low-$\Omega$ limit of
the factor $f(\Omega)$. Hence,
\begin{equation}
\nabla \cdot \bfv = - H_0 f(\Omega)\, \delta \,,
\label{eq:div_v_fin}
\end{equation}
in agreement with the general linear theory prediction, Equation~(\ref{eq:linear}). 

Now, we claim that in the limit $\Omega \to 0$,
solution~(\ref{eq:solut_limit}) is also a solution of the general
equation of motion~(\ref{eq:motion}), {\em for arbitrary\/}
$\delta$. To prove this statement we have to demonstrate that in this
limit, the non-linear term in equation~(\ref{eq:motion}) is
negligible. Substituting solution~(\ref{eq:solut_limit}) in this term
gives
\begin{equation}
\frac{\partial \bfv}{\partial t} + \frac{\dot{a}}{a} \bfv = \bfg -
\frac{1}{a} \left(H_0^{-1}\eta a \bfg \cdot \nabla\right)
(H_0^{-1}\eta a \bfg) \,.
\label{eq:motion_approx}
\end{equation}
The amplitude of the second term on the RHS is of order $H_0^{-2} \eta^2
a\, g\, \nabla \cdot \bfg \sim H_0^{-2} \eta^2 a\, g\, H^2 \Omega a\,
\delta$. The amplitude of the second term relative to the first is
thus 
\begin{equation}
\frac{2\mathrm{nd}}{1\mathrm{st}} \sim H_0^{-2} (H a)^2 \eta^2 \Omega
\, \delta \sim \eta^2 \Omega \,\delta \sim \eta^2 {\rm e}^{-\eta} \delta
\,, 
\label{eq:ratio}
\end{equation}
and in the limit $\Omega \to 0$ it tends to zero. (Formally speaking,
for arbitrary $\eps > 0$ and arbitrary $\delta$, there always exists
$\eta_\eps$ such that for all $\eta > \eta_\eps$, $\eta^2 {\rm
  e}^{-\eta} |\delta| < \eps$.) Thus, in the limit $\Omega \to 0$ the
non-linear term in the equation of motion becomes negligible, for
arbitrary value of $\delta$. This is why in every matter-only, open
universe, the velocity--density relation evolves towards the linear
one.

\section{Third-order expansion for $\bmath{\Theta}$ in voids}
\label{app:Theta3} 
Our aim here is to extend calculations of Section~\ref{sec:voids} for
voids up to third order in the perturbation parameter $\phi$ ($\phi$
is assumed to be large, but not infinitely large). We begin applying
to Equation~(\ref{form:delta fi eta}) the equality $\sinh\phi =
\cosh\phi\, -\, \exp(-\phi)$ and expand this equation up to terms of the
order $\cosh^{-2}\phi$. Solving perturbatively the resulting equation
for $\phi_3$ we obtain

\begin{equation}
\cosh\phi_3 = \cosh\phi_2 - \frac{3\phi_1^2 - 10\phi_1 +
  10}{\cosh\phi_1} \,,
\label{eq:phi3}
\end{equation}
where $\cosh\phi_2$ is given by Equation~(\ref{eq:phi2_eta_arbitrary}),
$\phi_1$ by~(\ref{eq:phi1_eta_arbitrary}) and the second term on the
RHS of the above equation is a small correction. This enables us to
write 

\begin{equation}
\sqrt{1 + \cosh\phi_3} \simeq \sqrt{1 + \cosh\phi_2} - \frac{3\phi_1^2
  - 10\phi_1 + 10}{2\, \cosh^{3/2}\phi_1} \,.
\label{eq:sqrt_phi3}
\end{equation}
Using the above equation in Equation~(\ref{form:teta open}) yields
\begin{eqnarray}
\Theta_3 = \Theta_2 + \frac{3}{2} \sqrt{\frac{\Omega}{2}}
\frac{(\sinh\eta - \eta)^3}{(\cosh\eta - 1)^{9/2}} \left(3\phi_1^2
  - 10\phi_1 + 10 \right) \times && \nonumber \\
(1 + \delta)^2 , ~ &&
\label{eq:Theta3}
\end{eqnarray}
or, finally,

\begin{equation}
\Theta_3 = \Theta_2 + 
\frac{3\,(\sinh\eta - \eta)^3}{2\,\sinh\eta\, (\cosh\eta - 1)^4}
\, F(\delta,\eta)\, (1 + \delta)^2 \,.
\label{eq:Theta3_fin}
\end{equation}
Here, 
\begin{eqnarray}
F(\delta,\eta) = 3\ln^2(1+\delta) \!\!\!\!\!&+&\!\!\!\!\! [10-6\ln(2g)]\,
\ln(1+\delta) + 3\ln^2(2g) \nonumber \\
\!\!\!\!\!&-&\!\!\!\!\! 10 \ln(2g) + 10 \,,
\label{eq:F_delta_eta}
\end{eqnarray}
and $g(\eta)$ is given by Equation~(\ref{eq:g_eta}). Inspecting terms
in the above equation we see that Equation~(\ref{eq:Theta3_fin}) can
be indeed written in the form~(\ref{eq:Theta3_general}). 

\end{document}